\documentclass[12pt]{article}
\usepackage{graphicx}
\usepackage{palatino}
\begin{document}
\def \non{\nonumber}
\def \ra{\rightarrow}
\def \bea{\begin{eqnarray}}
\def \eea{\end{eqnarray}}
\def \Qbar{\overline Q}
\def \qqbar{Q\Qbar}
\def \slj{^{2S+1}L_J}
\def \jpsi{J/\psi}
\def \ppbar{p\overline{p}}
\def \ccbar{c\overline{c}}
\def \pt{$p_T$}
\def \sg{\sigma}
\begin{titlepage}
\phantom{TIFR}
\bigskip
\begin{flushright}\vbox{\begin{tabular}{l}
	   IMSc/2002/12/43 \\
           TIFR/TH/02-36\\
           December, 2002\\
\end{tabular}}\end{flushright}
\begin{center}
{\Large \bf \boldmath 
$\eta_c$ production and dimuon enhancement \\in heavy ion collisions}
\end{center}
\bigskip
\begin{center}
   {{\bf Rahul Basu}\footnote{E-mail: rahul@imsc.res.in}\\
    {\sl The Institute of Mathematical Sciences,\\
    C.I.T. Campus, Chennai 600 113, India.}\\

\vskip12pt 
    {\bf K.~Sridhar}\footnote{E-mail: sridhar@theory.tifr.res.in}\\

    {\sl Department of Theoretical Physics,\\
     Tata Institute of Fundamental Research,\\
    Homi Bhabha Road, Mumbai 400005, India.}}
\end{center}
\bigskip
\begin{abstract}
\noindent Dilepton production in heavy ion collisions, in the
Intermediate Mass Region (IMR) has consistently shown an excess over
theoretical estimates. An attempt to understand this discrepancy between 
the observed
dilepton pairs and the theoretical estimate is made here through
the production of the $\eta_c$ meson and estimates obtained by
NRQCD calculations. We find that $\eta_c$ production offers a
satisfactory quantitative picture for explaining the discrepancy.
\end{abstract}
\end{titlepage}
Dilepton production plays a very important role in the study and understanding
of heavy ion collisions. This is mainly because dileptons do not interact
with the surrounding hadronic medium after being produced in a 
nucleus-nucleus collision. The typical dilepton invariant mass spectrum 
appears as
a wide continuum interrupted by various resonance ($\rho,\omega,\phi,
J/\psi,\psi^\prime$ etc) decay peaks. 

The Intermediate Mass Region (between the $\phi$ and $J/\psi$ peaks, 1.5 GeV 
to 2.5 GeV) is particularly interesting because it is believed to contain 
dileptons created in the thermalised QGP produced in nucleus-nucleus 
collisions (thermal dimuons). However it is precisely in this region that 
many different experiments in A-A collisions \cite{exp} have shown an excess
of dimuon production. In all these data sets, the dilepton sources are either 
Drell-Yan pairs or decays of $J/\psi$ or $D\bar D$. While the bulk of the data
agrees with such a production picture, there is a significant discrepancy 
between the observed dilepton pairs and the theoretical estimate based on the 
above sources in this Intermediate Mass region (IMR) with $\mu^+\mu^-$ 
invariant mass in the range 1.5 -- 2.5 GeV.

Many explanations have been offered in the literature for this excess {\it viz.}
decrease in $\rho$ meson mass due to thermal effects in $e^+e^-$ data 
\cite{lkb}, D-rescattering \cite{lw},
enhanced $D{\bar D}$ production, in-flight $\pi^+\pi^-$ decaying to 
$e^+e^-$ \cite{tserruya}, fireball hydrodynamics \cite{rs} and so on. 

In the entire kinematic range (upto 5 GeV) other than the IMR region 
mentioned above, where theory and experiment agree, the overall picture 
involving charm quarks \cite{kharzeev} is that when
there is sufficient energy exchange in a
collision, protons have non-negligible charm content ($c{\bar c}$
pairs), and substantial high
energy gluons which in turn can decay to $c{\bar c}$ pairs. These
$c{\bar c}$ pairs occassionally form bound states such as $J/\psi$ by 
emitting a soft gluon to maintain color balance, or can further polarize 
$u{\bar u}$ or $d{\bar d}$
from the surrounding medium to form $D{\bar D}$ pairs. Although the present
theoretical understanding cannot predict absolute numbers for these
processes it is possible to check the consistency of this picture with
various p-A and A-A data. By and large the data agrees with various 
quantitative checks.

Other charm meson bound states can also be produced, such as $\eta_c,
\psi^\prime$ and $\chi^\prime$'s, though their relative abundance is 
constrained by size, the larger ones being less likely to be formed. 

However the $\eta_c$ meson, which is a 1S orbital state is expected to
have the same size and has almost the same mass as the $J/\psi$. They only 
differ in their spin and hence in
any collision where $c{\bar c}$ quarks are produced these can form $\eta_c$
with about $1/3$ probability as $J/\psi$. However the possibility of the 
production of $\eta_c$ mesons in nucleus-nucleus collisions was only 
considered very recently in \cite{ab} wherein this was presented as a 
possible explanation of the discrepancy between theory and experiment 
in nucleus-nucleus collision -- in particular in S-U and Pb-Pb collisions. 
However this paper presented a somewhat qualitatitive field theoretical 
picture of the production of $\eta_c$ mesons and used it to explain the 
discrepancy in the IMR region. Furthermore, this simple field theoretical
picture was unable to account for the discrepancy seen in the IMR region 
for Pb-Pb collisions in the central region. 

In what follows we attempt to understand the discrepancy in the IMR region
once again through the production of the $\eta_c$ but by using a more 
quantitative description for the production of the $\eta_c$ meson. Our 
attempt may, therefore, be seen as a way to build upon the ideas of \cite{ab}
using the technology of NRQCD to present a more quantitative picture. 

In NRQCD \cite{BBL, kramer}, the quarkonium production 
cross-section factorises into a perturbatively calculable 
short-distance ($\leq 1/M_Q$, $M_Q$ is the mass of the heavy quark) 
effect and a long-distance part which is given by non-perturbative
matrix elements. 
The cross section for production of a quarkonium state $H$ can be 
written as
\bea
\sg(H)=\sum_n\frac{F_n}{M_Q^{d_n-4}}\,\left<0\left|{\cal O}_n^H
	\right|0\right>.
\label{factorizn}
\eea
The coefficients $F_n$ correspond to the production of $\qqbar$ in the 
angular momentum and colour state (singlet or octet) denoted by $n$ 
and is calculated using perturbative QCD.  The non-perturbative part, 
$\langle{\cal O}_n^H 
\rangle$ of mass dimension $d_n$ in NRQCD has a well-defined operator 
definition and is universal. These matrix elements can be extracted from 
any one process and then be used to predict other processes where 
the same matrix elements appear. Though the summation involves an infinite
number of terms, the relative magnitude of the various terms is
predicted by NRQCD and these matrix elements scale as powers of the
relative velocity $v$. 
However, this does not necessarily imply that effects from higher
orders in $v$ will always be small in physical processes, because
any observable, like the decay width or the cross
section is given by a double expansion in the strong coupling constant 
$\alpha_s(M_Q)$ 
and the relative velocity $v$. 

The non-perturbative matrix elements in NRQCD, are not calculable 
and have to be obtained by fitting to available data. The matrix 
elements of the colour-singlet operators 
can be obtained from the quarkonium decay widths 
and the colour-octet matrix elements 
have been obtained by fitting NRQCD predictions to the
CDF data \cite{cho1, cho2}. 
The remarkable thing is that the non-perturbative
parameters appearing in the $\eta_c$ production cross section can be
determined from the matrix elements determined from $J/\psi$ production
at the Tevatron: this happens because of the heavy-quark symmetry of
the NRQCD Lagrangian. This has been exploited earlier in the context
of $h_c$ and $\eta_c$ production at Tevatron \cite{sri, sri2}. 

A Fock space expansion of the physical $\eta_c$, which is a $^1S_0$
($J^{PC}=0^{-+}$) state yields
\bea
\!  \!  \!  \!
\left|\eta_c\right>={\cal O}(1)
	\,\left|\qqbar[^1S_0^{[1]}] \right>+
	 {\cal O}(v^2)\,\left|\qqbar[^1P_1^{[8]}]\,g \right>+
	{\cal O}(v^4)\,\left|\qqbar[^3S_1^{[8]}]\,g \right>+\cdots ~.
\label{fockexpn}
\eea
The colour-singlet $^1S_0$ state contributes at ${\cal O}(1)$ but the
colour-octet $^1P_1$ and $^3S_1$ channels effectively contribute
at the same order because the $P$-state production is itself down
by a factor of ${\cal O}(v^2)$.
The colour-octet states become a physical $\eta_c$ by the $^1P_1^{[8]}$ 
state emitting a gluon in an E1 transition, and by the $^3S_1^{[8]}$
state emitting a gluon in an M1 transition.  
The contributing subprocess cross sections are
\bea
q ~\bar q &\ra& \qqbar[\slj] , \non\\
g ~g &\ra& \qqbar[\slj] ,\non
\eea
where the $\qqbar$ is in the $^1 S^{[1]}_0$, $^1 S^{[8]}_0$ and 
$^3 S^{[8]}_1$ states.
The $^1 P^{[8]}_1$ state does not contribute at this leading order 
in $\alpha_s$ because of Yang's theorem.

We have computed the contributions to the cross-section for $\eta_c$
production from the $^1 S^{[1]}_0$, $^1 S^{[8]}_0$ and $^3 S^{[8]}_1$ states. 

Heavy quark spin-symmetry is made use of in obtaining 
$\left< {\cal O}_n^{\eta_c}\right>$'s from the experimentally available
$\langle {\cal O}_n^{J/\psi}\rangle$'s. Using this symmetry we get the 
following relations among $\langle {\cal O}_n^H \rangle$'s:
\bea
\left< 0 \right|{\cal O}_1^{\eta_c}[^1S_0]\left| 0\right>&=&
\left< 0 \right|{\cal O}_1^{J/\psi}[^3S_1]\left| 0 \right>
\;(1+ O(v^2)), \non\\
\left< 0 \right|{\cal O}_8^{\eta_c}[^3S_1]\left| 0\right>&=&
\left< 0 \right|{\cal O}_8^{J/\psi}[^1S_0]\left| 0 \right>
\;(1+ O(v^2)).
\label{Ovalues}
\eea
For the singlet matrix elements we have $\left< 0 
\right|{\cal O}_1^{J/\psi}[^3S_1]\left| 0 \right>=1.2$ GeV$^3$ 
The CDF $J/\psi$ data only constrains a combination of
octet matrix elements given by $A_1+A_2 \equiv$
\(\frac{\left< 0 \right|{\cal O}_8^{J/\psi}[^3P_0]\left| 0 \right>}
{M_c^2}+
\frac{\left< 0 \right|{\cal O}_8^{J/\psi}[^1S_0]\left| 0 \right>}{3}
=(2.2\pm0.5)\times 10^{-2}\) GeV$^3$ \cite{cho2}. The CDF $J/\psi$ data
do not allow for a separate determination of the values of $A_1$ and
$A_2$ because the shapes of these two contributions to the $J/\psi$ $p_T$
distribution are almost identical. For our
numerical predictions we assume that the non-perturbative matrix element
of interest to us ($\left< 0 \right|{\cal O}_8^{\eta_c}[^3S_1]\left| 
0\right>$) lies in the range determined by this sum.
\begin{figure}[htb]
\begin{center}
\includegraphics[scale=0.35]{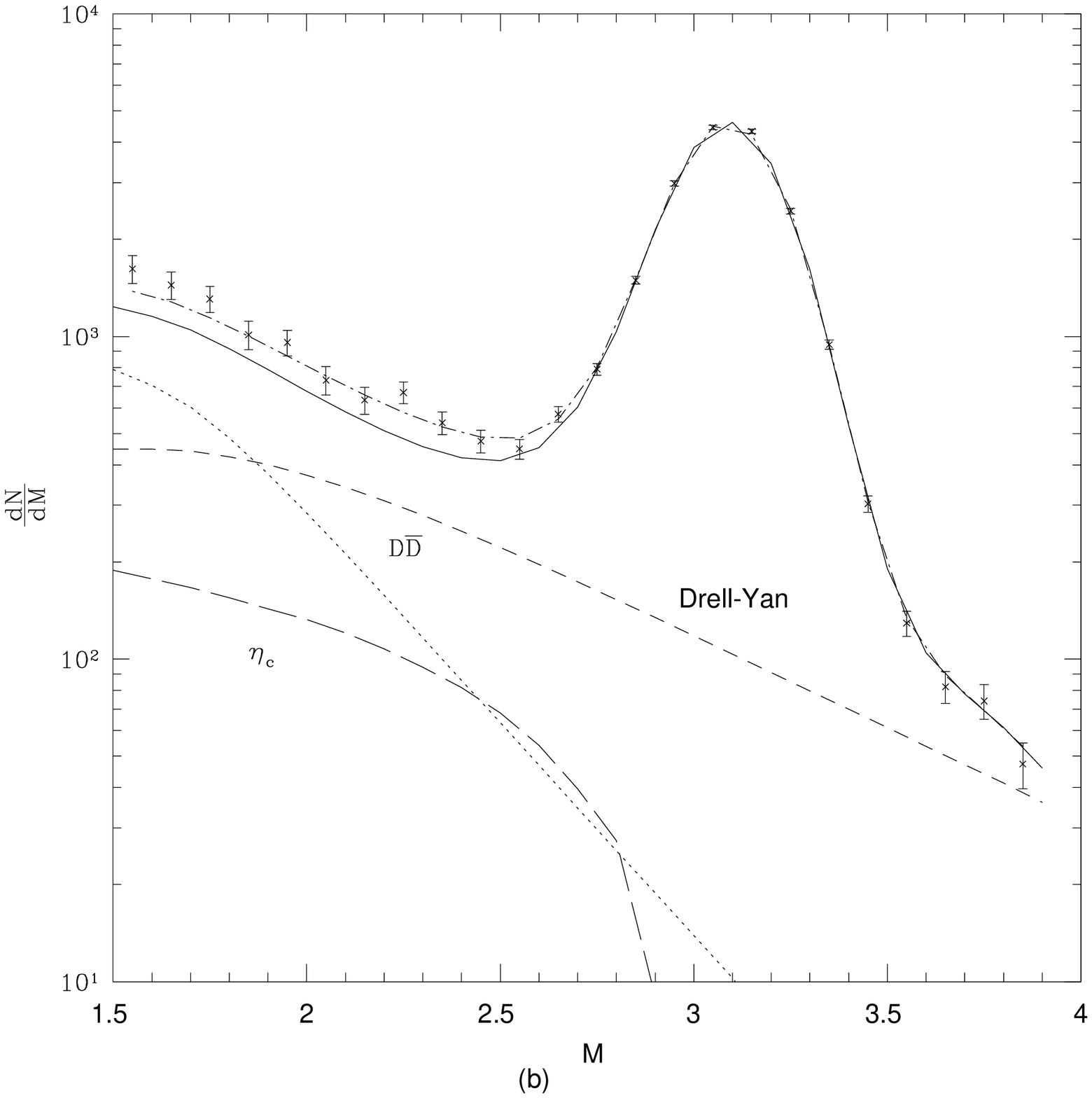}\hfil \includegraphics[scale=0.35]{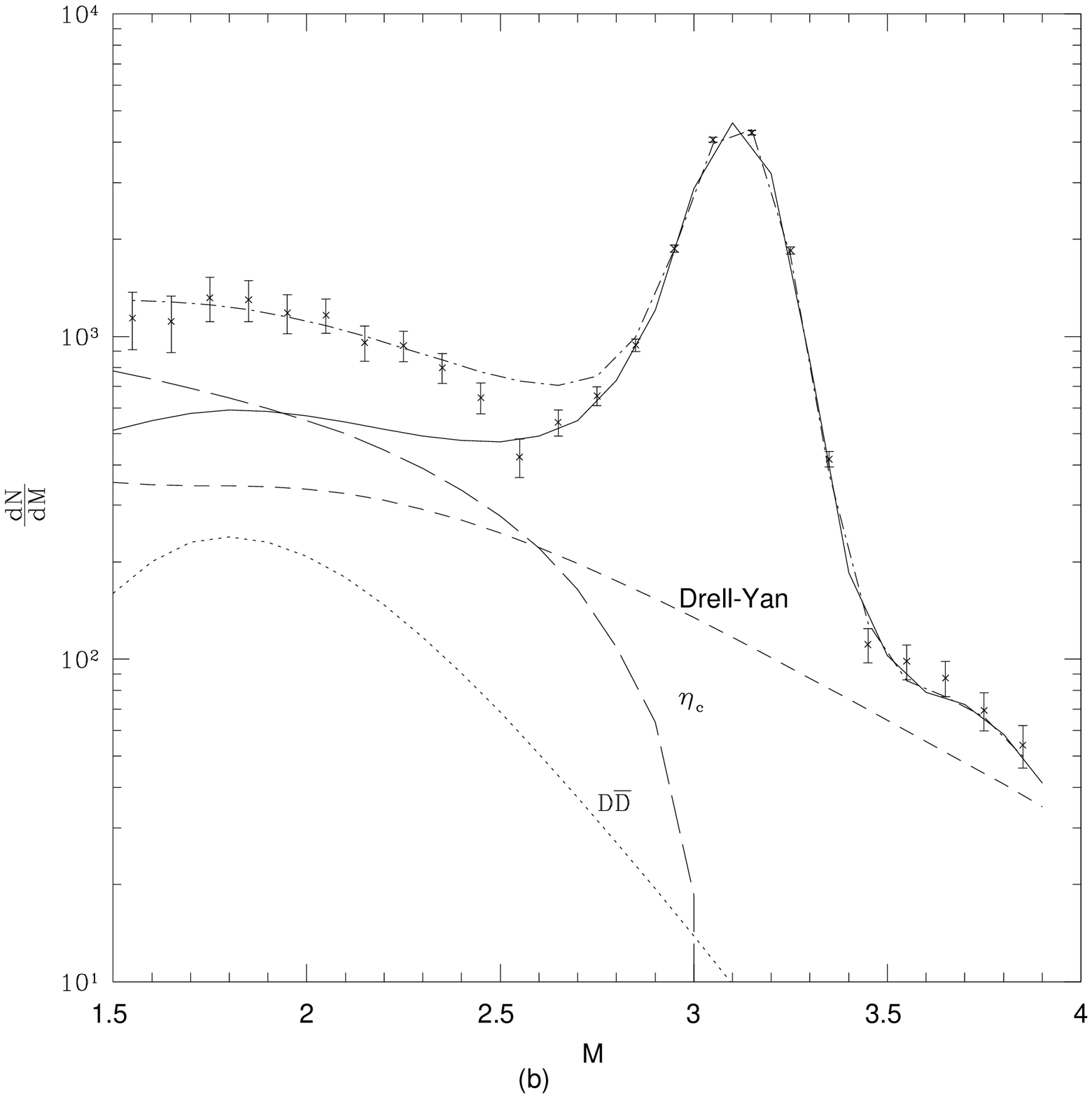}
\end{center}
\caption{The graphs show the S-U (a) and Pb-Pb (b) results for 
central collisions.
Note that in the IMR region the $\eta_c$ contribution appears to
saturate the deficit.}
\end{figure}

Using these values for the non-perturbative matrix elements we can compute
the cross-section for $\eta_c$ production. This has to be convoluted with
the branching ratio for the $\eta_c \rightarrow \gamma \gamma^* \rightarrow
\gamma \mu^+ \mu^-$. This quantity can be estimated only approximately and
to do this we have used the value of the $\eta_c \rightarrow \gamma \gamma$ 
branching ratio modulated by a $(1-M_{\gamma^*}^2/M_{\eta_c}^2)$ in the
denominator. With these inputs, we have computed the $\mu^+ \mu^- \gamma$
yield in Pb-Pb and S-U collisions. The nuclear parton distributions have
been taken from the parameterisation of EKS \cite{eks}. The cuts and
acceptances for the muons and photons have been taken from the experimental
papers (see, for example, \cite{na38}). 

Our results for $dN/dM$ are compared to the experimental curves for
central collisions in in Fig. 1. In keeping with the convention used by
the experimentalists, we have indicated the $\eta_c$ production
contribution separately (long-dashed line) as well as the summed contribution 
of all the individual contributions in the IMR region (dashed-dotted
line). It appears from both the graphs that the $\eta_c$ contribution
saturates the discrepancy seen in the IMR region. However this agreement
is based on certain assumptions. First of all, the branching ratio of
$\eta_c \rightarrow \gamma \gamma^*\rightarrow \gamma\mu^+\mu^-$ used, 
as already stated in the previous paragraph, is only
approximately estimated through very general field theoretical
arguments. Secondly, as with all leading order calculations, the result
is dependent (though mildly) on the scale of $\alpha_s$ used in the
calculation. Finally since, as we have explained, the CDF $J/\psi$ data
only constrains the combination of octet matrix elements given by $A_1$ and
$A_2$ we have, for convenience, assumed that the non-perturbative
matrix element $<0|{\cal O}^{J/\psi}_8[^1S_0]|0>$ (i.e $A_2$) saturates the sum 
and hence gives the value of the $<0|{\cal O}^{\eta_c}_8[^3S_1]|0>$
matrix element. However this turns out to be not such a serious
approximation and we find that varying this value between $A_1$ and
$A_2$ produces very little change in the overall shape and magnitude of
the $\eta_c$ contribution. 

The general arguments for the $\eta_c$ contribution used in \cite{ab} 
were unable to explain 
the discrepancy in central collision in Pb-Pb collision and it was 
speculated there that the difference  between theory and experiment even
after including $\eta_c$ production, could perhaps be accounted for by
glueball production.  It seems however, in the present, more
quantitative, analysis, that this possibility is ruled out since the
dileptons from $\eta_c$ seem to cover the deficit. In any case, even if
such a scenario of glueball production is envisaged, the contribution
would clearly be very small.

Finally we have made no comments about peripheral collisions where also
this discrepancy in the IMR region is seen. This is
because a quantitative study of peripheral collision would involve a
detailed understanding of the geometry of the collision, and its
consequent uncertainties. In view of these, we feel that no firm
statements can really be made within the context of this approach, to
peripheral collisions in heavy ion collisions.

In conclusion, in this paper, we have used the technology of NRQCD
to estimate the contribution of the $\eta_c$ meson to dilepton
production in the IMR region of heavy ion collisions, and found that it
is able to satisfactorily explain the deficit, within the limits of the
assumptions used in this study.


\newpage 

\begin{thebibliography}{99}
\bibitem{exp}
I. Ravinovich for CERES Coll., Nucl. Phys. {\bf A638}, 159C (1998);
G. Agakichiev for CERES Coll., Nucl. Phys. {\bf B422}, 405 (1998);
M. Masera for HELIOS3 Coll., Nucl. Phys. {\bf A590}, 93C (1995);
A. De Falco for NA38 Coll., Nucl. Phys. {\bf A638}, 487C (1998);
E. Scomparin for NA50 Coll., J. Phys. {\bf G25}, 235 (1999);
A. Drees, Nucl. Phys. {\bf A610}, 536C (1996).
\bibitem{lkb}
G. Q. Li, C. M. Ko and G. E. Brown, Phys. Rev. Lett. {\bf 75}, 4007
(1995); Nucl. Phys. {\bf A606}, 568 (1996).
\bibitem{lw}
Z. Lin and X. N. Wang, Phys. Lett. {\bf B444}, 245 (1998).
\bibitem{tserruya}
I. Tserruya, LANL Archives nucl-ex/9912003 and references therein.
\bibitem{rs}
R. Rapp and E. Shuryak, Phys. Lett. {\bf B473}, 13 (2000).
\bibitem{kharzeev}
D. Kharzeev, Nucl. Physics {\bf A638}, 279C (1998) and references
therein.
\bibitem{ab}
R. Anishetty and R. Basu, Phys. Lett. {\bf B495}, 295 (2000).
\bibitem{BBL} G.T.~Bodwin, E.~Braaten and G.P.~Lepage, Phys. Rev. 
{\bf D51}, 1125 (1995);erratum $ibid.$ {\bf 55} 5853 (1997).
\bibitem{kramer} M. Kr\" amer, Prog. Part. Nucl. Phys. 47 141 (2001). 
\bibitem{cho1} P. Cho and A.K.~Leibovich,  Phys. Rev.  {\bf D53}, 
150 (1996).
\bibitem{cho2} P. Cho and A.K.~Leibovich,  Phys. Rev.  {\bf D53},
6203 (1996).
\bibitem{sri} K. Sridhar,  Phys. Rev. Lett., {\bf 77}, 4880 (1996).
\bibitem{sri2} P. Mathews, P. Poulose and K. Sridhar, Phys. Lett. B438 336 
(1998).
\bibitem{eks} K.J. Eskola, V.J. Kolhinen and P.V. Ruuskanen,
Nucl. Phys. B 535 (1998) 351.
\bibitem{na38}M. C. Abreu et al., CERN-EP-2000-012,  Eur. Phys. J.  
{\bf C14}, 443 (2000).
\end{thebibliography}
\end{document}